\documentclass[a4paper,fleqn,usenatbib,useAMS]{mnras}

\usepackage{float}
\usepackage{graphicx}
\usepackage{times}
\usepackage{amstext}
\usepackage{amsmath}
\usepackage{amssymb}	% Extra maths symbols
\usepackage{natbib}
\usepackage{url}
\usepackage{tabularx}
\usepackage{mathtools}
\usepackage{mathrsfs}
\usepackage{graphics}
\usepackage{multirow}
\usepackage{ar}

\newcommand{\kms}{\,km\,s$^{-1}$} % kilometres per second
\newcommand{\solar}{M$_{\odot}$}
\newcommand{\irdc}{${\rm G}035.39-00.33$}

\usepackage[T1]{fontenc}
\usepackage{ae,aecompl}

\usepackage{mathptmx}
\usepackage{txfonts}

\title[The early-stage anatomy of a protocluster hub]{Unveiling the early-stage anatomy of a protocluster hub with ALMA}

\author[J. D. Henshaw et al.]{J. D. Henshaw$^{1,2}$ \thanks{Contact e-mail: j.d.henshaw@ljmu.ac.uk}, I. Jim\'{e}nez-Serra$^{3,4}$, S. N. Longmore$^{1}$, P. Caselli$^{5}$, J. E. Pineda$^{5}$, 
\newauthor A. Avison$^{6,7}$, A. T. Barnes$^{1,5}$, J. C. Tan$^{8}$, and F. Fontani$^{9}$   \\
$^{1}$ Astrophysics Research Institute, Liverpool John Moores University, Liverpool, L3 5RF, UK\\
$^{2}$ School of Physics and Astronomy, University of Leeds, Leeds LS2 9JT, UK\\
$^{3}$ School of Physics and Astronomy, Queen Mary University of London, Mile End Road, London E1 4NS\\
$^{4}$ University College London, 132 Hampstead Road, London, NW1 2PS, UK\\
$^{5}$ Max-Planck Institute for Extraterrestrial Physics, Giessenbachstrasse 1, 85748 Garching, Germany\\
$^{6}$ Jodrell Bank Centre for Astrophysics, School of Physics and Astronomy, The University of Manchester, Manchester, M13 9PL, UK\\
$^{7}$ UK ALMA Regional Centre Node, Manchester, M13 9PL, UK \\
$^{8}$ Departments of Astronomy \& Physics, University of Florida, Gainesville, FL, 32611, USA \\
$^{9}$ INAF-Osservatorio Astrofisico di Arcetri, Largo E. Fermi 5, I-50125 Firenze, Italy
}
\date{Accepted 2016 July 28. Received 2016 July 27; in original form 2016 June 27}

\pubyear{2016}

\begin{document}
\label{firstpage}
\pagerange{\pageref{firstpage}--\pageref{lastpage}}
\maketitle

% Abstract of the paper
\begin{abstract}
High-mass stars shape the interstellar medium in galaxies, and yet, largely because the initial conditions are poorly constrained, we do not know how they form. One possibility is that high-mass stars and star clusters form at the junction of filamentary networks, referred to as `hubs'. In this Letter we present the complex anatomy of a protocluster hub within an Infrared Dark Cloud (IRDC), \irdc, believed to be in an early phase of its evolution. We use high-angular resolution ($\{\theta_{\rm maj},\,\theta_{\rm min}\}=\{1.4\,{\rm arcsec},\,0.8\,{\rm arcsec}\}\sim\{0.02\,{\rm pc},\,0.01\,{\rm pc}\}$) and high-sensitivity ($0.2$\,mJy\,beam$^{-1}$; $\sim0.2$\,\solar) 1.07\,mm dust continuum observations from the Atacama Large Millimeter Array (ALMA) to identify a network of narrow, $0.028\,\pm\,0.005$\,pc wide, filamentary structures. These are a factor of $\gtrsim3$ narrower than the proposed `quasi-universal' $\sim0.1$\,pc width of interstellar filaments. Additionally, 28 compact objects are reported, spanning a mass range $0.3\,{\rm M_{\odot}}<M_{\rm c}<10.4\,{\rm M_{\odot}}$. This indicates that at least some low-mass objects are forming coevally with more massive counterparts. Comparing to the popular `bead-on-a-string' analogy, the protocluster hub is poorly represented by a monolithic clump embedded within a single filament. Instead, it comprises multiple intra-hub filaments, each of which retains its integrity as an independent structure and possesses its own embedded core population. 
\end{abstract}

% Select between one and six entries from the list of approved keywords.
% Don't make up new ones.

%%%%%%%%%%%%%%%%%%%%%%%%%%%%%%%%%%%%%%%%%%%%%%%%%%

%%%%%%%%%%%%%%%%% BODY OF PAPER %%%%%%%%%%%%%%%%%%

\section{Introduction}\label{Section:introduction}

In recent years, Infrared Dark Clouds (IRDCs; \citealp{perault_1996, egan_1998}) have received significant attention in the field of high-mass ($>~8$\,\solar) star formation, owing to their natal association with massive ($\sim100$\,\solar) protocluster clumps. Early studies earmarked IRDC clumps as potential locations of high-mass star and star cluster formation (e.g. \citealp{rathborne_2006}), with high-angular resolution observations adding credence to this theory (\citealp{tan_2013b, tan_2016, cyganowski_2014, zhang_2015}). 

Common to many IRDCs is the presence of filamentary structure. As a key ingredient of the molecular interstellar medium \citep{andre_2010, molinari_2010}, there has been a commensurate drive towards understanding the intrinsic nature of filaments. Physically, interstellar filaments are characterized by a ``quasi-universal'' width of the order $\sim0.1$\,pc \citep{arzoumanian_2011}. The origin of this commonly observed width is currently unknown. One possibility is that this scale is a direct reflection of the physics of filament formation, following the dissipation of turbulent energy via shocks at the stagnation points of a turbulent velocity field (e.g. \citealp{federrath_2016}). From a dynamical perspective, both observations (e.g. \citealp{schneider_2012}) and simulations (\citealp{dale_2011, amyers_2013, gomez_2014}) agree that filaments play an important role in star and cluster formation. In this context, filaments may act as tributaries, funnelling mass towards protocluster clumps at the centre of so-called ``hub-filament'' systems \citep{myers_2009, liu_2012, gomez_2014, smith_2016}.

Observational studies of relatively evolved massive star-forming clusters support the ``hub-filament'' scenario for high-mass star and star cluster formation (e.g. \citealp{liu_2012}). For objects at an early stage of evolution, such as IRDCs, there is growing evidence in the form of coherent, parsec-scale, velocity gradients associated with prominent filaments in the vicinity of protocluster clumps (\citealp{peretto_2014, tackenberg_2014}). Although high-angular resolution observations of IRDC clumps exist, these often focus on the embedded core population (e.g. \citealp{wang_2014, beuther_2015b}), and typically lack enough spatial coverage and dynamic range to obtain a global understanding of the anatomy of hub-filament systems. As a consequence, the physical structure of protocluster hubs during the earliest phases of evolution is poorly documented. It remains unknown, for example, whether filamentary structures exist internally within hubs, and if so, what role they may play in the subsequent evolution of protocluster cores. 

In this Letter we present high-angular resolution ($\{\theta_{\rm maj},\,\theta_{\rm min}\}=\{1.4\,{\rm arcsec},\,0.8\,{\rm arcsec}\}$; see \S~\ref{Section:Observations}) and high-sensitivity (0.2 mJy\,beam$^{-1}$) Atacama Large Millimetre Array (ALMA) cycle 2 dust continuum observations towards a $\sim100$\,\solar \ clump \citep{rathborne_2006, butler_2012}, H6 (referred to as MM7 in \citealp{rathborne_2006}), located within the IRDC \irdc \ (cloud H; \citealp{butler_2012}). This cloud is reasonably massive ($\sim~2\times~10^{4}$\,\solar \ at a kinematic distance of $2.9\,{\rm kpc}$; \citealp{simon_2006b, kainulainen_2013}) and comprises a network of kinematically-identified sub-filaments that overlap spatially at the location of the H6 hub \citep{henshaw_2013, henshaw_2014, izaskun_2014}. Combining ALMA observations with those from the Atacama Compact Array (ACA), allows us to recover extended structures inside the hub, and unveil a network of narrow intra-hub filaments and embedded structure that had previously remained elusive. 

\begin{figure}
\begin{center}
\includegraphics[trim = 0mm 0mm -10mm 0mm, clip, width = 0.47\textwidth]{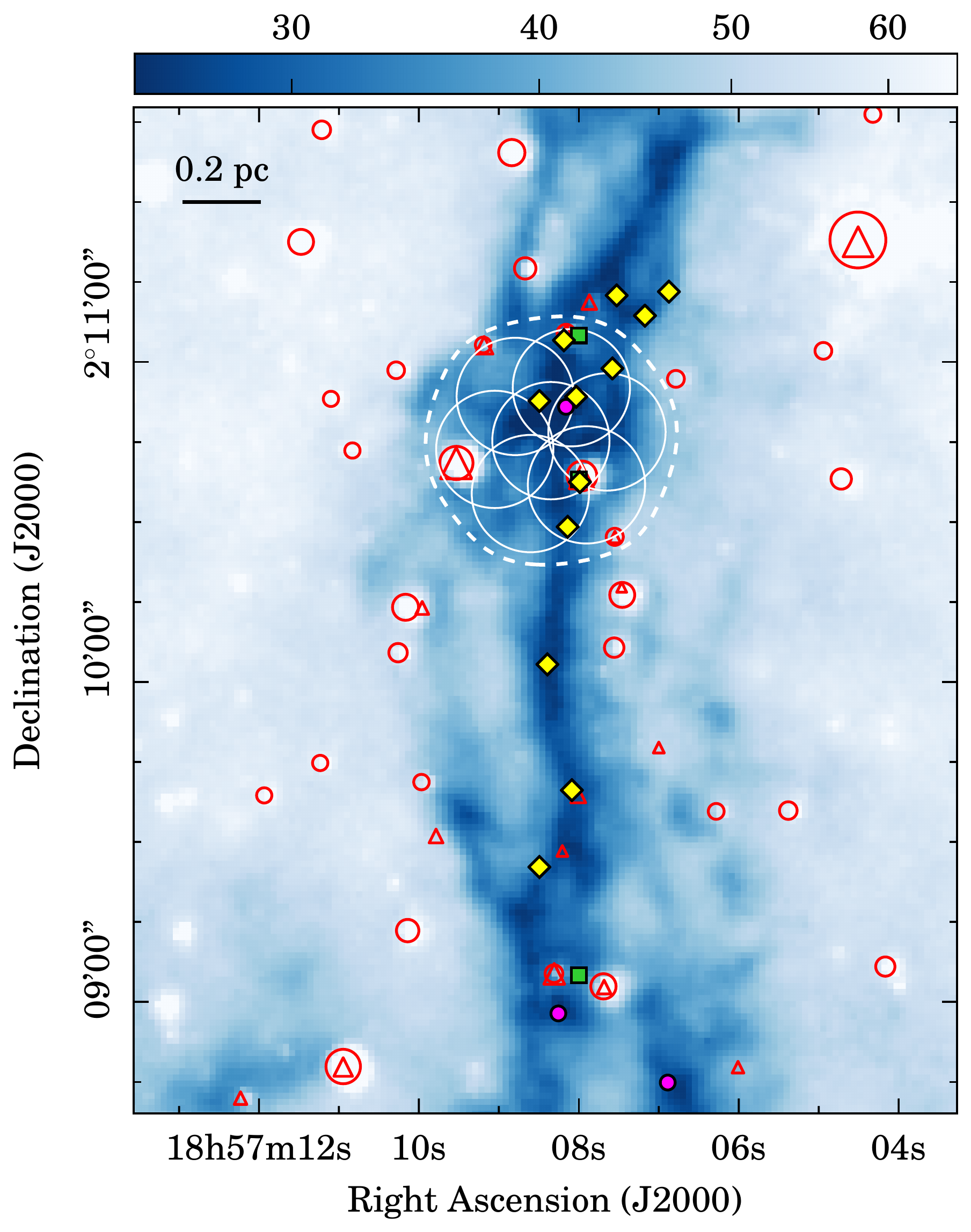}
\end{center}
\caption{\emph{Spitzer} 8\,\micron \ image of \irdc \ \citep{churchwell_2009}, with logarithmic intensity scale in units of MJy\,sr$^{-1}$. Open white circles indicate the 7-field 12\,m ALMA mosaic. The dashed contour displays the edge of the ALMA mosaic. Filled yellow diamonds and filled magenta circles indicate the locations of 3.2\,mm PdBI cores identified by \citet{henshaw_2016b} and high-mass cores reported by \citet{butler_2012}, respectively. Open red circles and red triangles refer to the 8\,\micron, and 24\,\micron\ emission, respectively \citep{carey_2009,izaskun_2010}. The symbol sizes are scaled by the source flux. Filled green squares highlight the location of extended 4.5\,\micron\ emission \citep{chambers_2009}.  }
\label{Figure:mirmap}
\end{figure}
\section{Observations}\label{Section:Observations}

ALMA cycle 2 Band 7 observations were carried out towards the H6 hub (project: 2013.1.01035.S, PI: Henshaw) using the 12~m ALMA and 7~m ACA arrays. The precipitable water vapour (PWV) was 1.53\,mm, resulting in system temperatures of $\sim130$\,K. Primary beam sizes of the 12~m and 7~m antennas at 279\,GHz are $\sim22$\,arcsec and 38\,arcsec, respectively. A 7-field mosaic was used to cover a region $\sim40\,{\rm arcsec}\times40\,{\rm arcsec}$ (corresponding to $0.55\,{\rm pc}\times0.55\,{\rm pc}$). The 12\,m mosaic can be seen as white open circles in Fig.~\ref{Figure:mirmap}, which shows the \emph{Spitzer} Infrared Array Camera (IRAC) 8\,\micron \ image of \irdc, from the Galactic Legacy Mid-Plane Survey Extraordinaire (GLIMPSE; \citealp{churchwell_2009}). The shortest available baseline from the 7~m array was $D=9$\,m. We are therefore sensitive to emission on scales up to $\sim30\,{\rm arcsec}$ ($\sim0.4\,{\rm pc})$. 
For the 12~m observations, quasars J1924-2914 and J1851+0035 were used for bandpass and time-dependent gain calibration (with measurements every $\sim$10~min), respectively. Flux calibration was performed using Neptune. The spectral set up included a 1.875\,GHz continuum band centred on 277.5\,GHz. The 12~m and 7~m visibility data were combined and imaged in {\sc casa}. Imaging and deconvolution was performed using MultiScale {\sc clean} with natural weighting. The resultant synthesized beam is $\{\theta_{\rm maj},\,\theta_{\rm min}\}=\{1.4\,{\rm arcsec},\,0.8\,{\rm arcsec}\}\sim\{0.02\,{\rm pc},\,0.01\,{\rm pc}\}\sim\{4000\,{\rm au},\,2000\,{\rm au}\}$ with a position angle ${\rm P.\,A.}=-57\fdg9$. The continuum image reaches a 1$\sigma_{\rm rms}$ noise level of 0.2\,mJy\,beam$^{-1}$ (estimated from emission-free regions after primary beam correction). From the \citet{rathborne_2006} 1.2\,mm dust continuum image of the H6 clump (size $\sim40$\,arcsec, $\sim$ to the extent of our ALMA mosaic), we estimate that the integrated flux at 1.07\,mm is $\sim0.8$\,Jy (assuming a dust emissivity index, $\beta=1.75$). The total flux in our ALMA image is $\sim0.5$\,Jy, indicating that we recover about 60 per cent of the total flux.

\section{Results}\label{Section:results}

\begin{figure*}
\begin{center}
\includegraphics[trim = 0mm 0mm 0mm 0mm, clip, width = 0.85\textwidth]{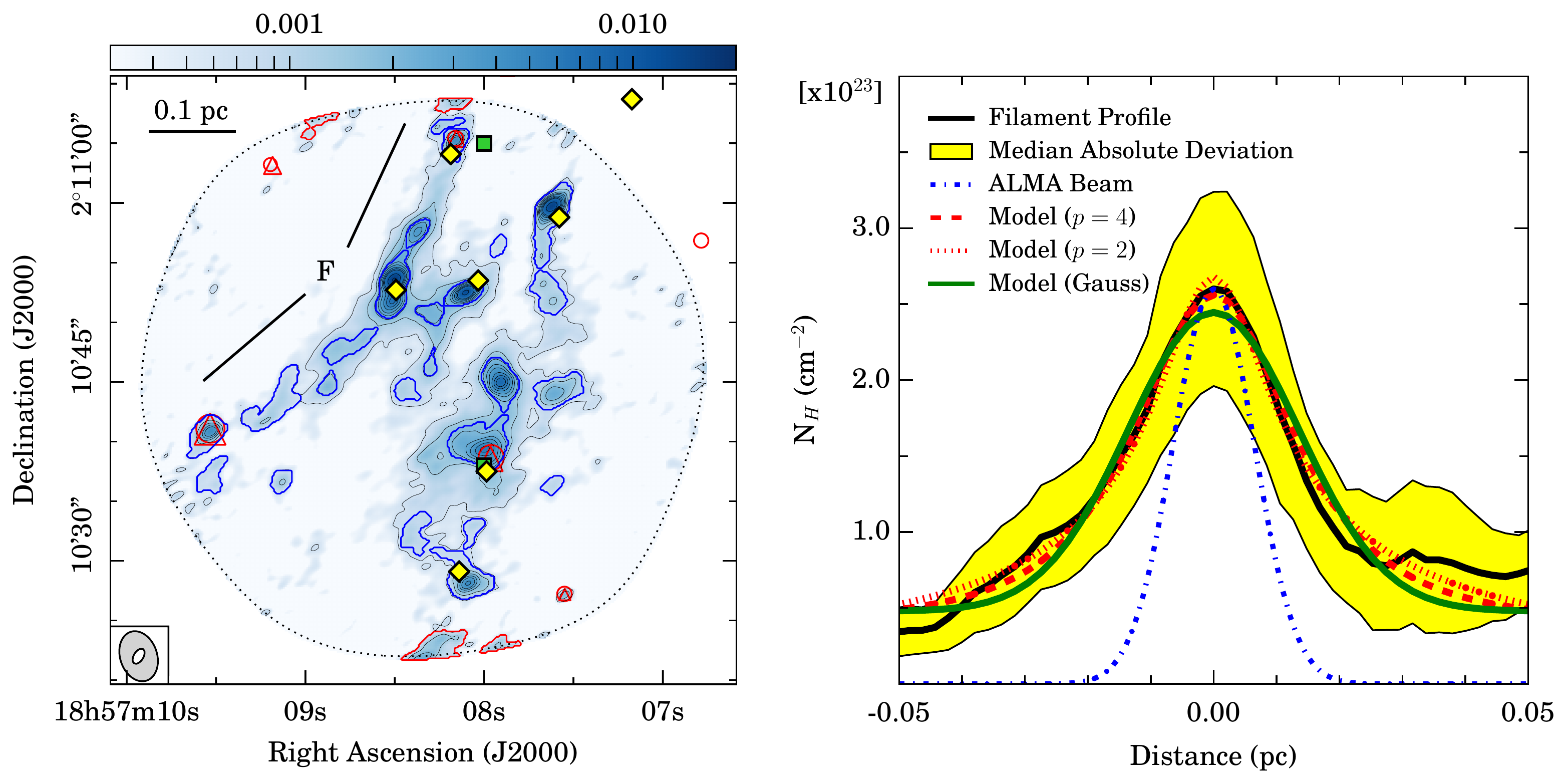}
\end{center}
\caption{Left: the ALMA 1.07\,mm continuum map (logarithmic colour scale in units of Jy\,beam$^{-1}$ and thin black contours). Contours increase in 3$\sigma_{\rm rms}$ ($\sigma_{\rm rms}\sim0.2$\,mJy\,beam$^{-1}$) steps from 3$\sigma_{\rm rms}$ to 15$\sigma_{\rm rms}$, then in 5$\sigma_{\rm rms}$ steps up until 30$\sigma_{\rm rms}$, before increasing in 20$\sigma_{\rm rms}$ steps until 70$\sigma_{\rm rms}$ ($\sim0.15$\,Jy\,beam$^{-1}$). The symbols and dotted contour have equivalent meaning to those presented in Fig.~\ref{Figure:mirmap}. The thick blue and red contours depict identified sources (\S~\ref{Section:results} and Table~\ref{Table:leaf_info}), with the latter representing leaves that are rejected from further analysis as they reside at the very edge of the map. The PdBI \citep{henshaw_2016b} and ALMA synthesized beams are shown in the bottom left-hand corner in grey and white, respectively. Right: The median column density profile perpendicular to the filament marked `F' in the left-hand panel (thick black line). The yellow area indicates the median absolute deviation. The ALMA synthesized beam is indicated by the blue line. Different models used to derive the filament width are shown in red and green (see \S~\ref{Section:results} for more details).  }
\label{Figure:cont_map}
\end{figure*}

The ALMA 1.07\,mm dust continuum emission map of the H6 region is displayed in the left-hand panel of Fig.~\ref{Figure:cont_map}. This new high-angular resolution and high-sensitivity map reveals a striking network of intra-hub filaments, which had remained elusive in previous observations (e.g. \citealp{henshaw_2016b}). 

The filamentary structures evident in Fig.~\ref{Figure:cont_map} are narrow in comparison to the prominent filament seen in extinction in Fig.~\ref{Figure:mirmap} and to the `quasi-universal' filament width of $\sim0.1$ pc \citep{arzoumanian_2011, andre_2016}. We can estimate the width of these filaments via the analysis of their radial surface density profiles, $\Sigma_{\rm f}(r)$. We focus on the filament marked `F' in Fig.~\ref{Figure:cont_map}, as it exhibits the most uniform structure of those observed. To derive the radial surface density profile we rotate the image and take horizontal slices of equivalent breadth across the filament crest. The column density, $N_{\rm H}$, is estimated using the following assumptions: (i) the dust continuum emission is optically thin; (ii) a total (gas plus dust)-to-dust-mass ratio, $R_{\rm gd}=141$ \citep{draine_2011}; (iii) a dust opacity per unit mass, $\kappa_{\nu}\approx1.26$\,cm$^{2}$g$^{-1}$ (valid for the moderately coagulated thin ice mantle dust model of \citealp{ossenkopf_1994}, assuming $\beta=1.75$); (iv) a dust temperature, $T_{\rm d}=13$\,K. For an in-depth discussion on these considerations, please see \citet{henshaw_2016b}. After accounting for uncertainties in our assumed parameters, the uncertainty in $N_{\rm H}$ is $\sim50$\,per cent. The resultant 1$\sigma_{\rm rms}$ sensitivity is $\sim4\times10^{22}\,{\rm cm}^{-2}$. 

We follow a conservative approach to estimating the filament width, using the median, rather than mean surface density profile (as is more commonly used). The presence of embedded cores means that the mean surface density profile typically yields narrower filaments. For completeness however, we derive the filament width using both profiles. The median radial surface density profile of the filament marked `F' in left-hand panel of Fig.~\ref{Figure:cont_map} is presented in the right-hand panel. 

Two independent models are used to establish the filament width (\citealp{smith_2014, federrath_2016}). First we consider the filament as an idealized cylinder, the surface density profile of which can be described by a Plummer profile of the form
\begin{equation}
\Sigma_{\rm f}(r) = A_{p}\frac{\Sigma_{\rm f,0}}{[1+(r/R_{\rm flat})^{2}]^{\frac{p-1}{2}}}+\Sigma_{\rm bg},
\label{Eq:plummer}
\end{equation}
where $r$ refers to the cylinder radius, $A_{p}$ is a finite constant factor dependent on the density profile power-law index, $p$, and the filament inclination angle, $i$ (for simplicity we assume $i=0$), $\Sigma_{\rm f,0}=\rho_{\rm f,0}R_{\rm flat}$, where $\rho_{\rm f,0}$ is the filament central density and $R_{\rm flat}$ is the radius of the flat inner section of the cylinder. For $p=2$, the filament width is $W_{\rm f}\sim3R_{\rm flat}$ \citep{arzoumanian_2011}. The second model involves fitting a simple Gaussian profile of the form
\begin{equation}
\Sigma_{\rm f}(r)=\Sigma_{\rm f, 0}{\rm exp}\,\bigg(-\frac{r^{2}}{2\sigma^{2}_{\rm Gauss}}\bigg)+\Sigma_{\rm bg},
\label{Eq:gauss}
\end{equation}
where $\sigma_{\rm Gauss}$ is the surface density dispersion about the filament centre and the width is given by $W_{\rm f}=2(2{\rm ln}\,2)^{1/2}\,\sigma_{\rm Gauss}$.

In Fig.~\ref{Figure:cont_map} we show the best fitting model solutions to the radial column density profile. The models described by equation~\ref{Eq:plummer} are shown in red. The dashed and dotted lines are the solutions assuming $p=4$ and $p=2$, respectively, where the former is the \citet{ostriker_1964} solution for an isothermal filament in hydrostatic equilibrium. The Gaussian model is shown in green. From Fig.~\ref{Figure:cont_map}, a Plummer-like surface density profile appears to better represent the observations, particularly within $\pm0.01$ pc and for radial distances $<-0.02$\,pc and $>0.02$\,pc. Taking $p=2$ (equation~\ref{Eq:plummer}), we find $R_{\rm flat}=0.009\pm0.001$\,pc. The resultant width is $W_{\rm f}\approx~0.027\,\pm\,0.006$\,pc. To investigate the possibility of variation in the filament width, we also fit individual profiles. From a sample of $\sim150$ radial cuts, we find a mean width of $W_{\rm f}\approx~0.029\,\pm\,0.013$\,pc (where the uncertainty reflects the standard deviation). Alternatively, the Gaussian model returns a full-width at half-maximum width, after beam deconvolution, of $W_{\rm f}\approx~0.028\,\pm~\,0.005$\,pc.\footnote{Using the mean column density profile gives widths of $W_{\rm f}~\approx0.021$\,pc or $0.020$\,pc for the Plummer and Gaussian models, respectively. } 

Our ALMA continuum image also reveals the presence of 28 high-contrast, compact cores embedded within the filaments. In Fig.~\ref{Figure:mirmap} and Fig.~\ref{Figure:cont_map} (left-hand panel), yellow diamonds indicate the position of the 3.2\,mm continuum sources previously identified with the Plateau de Bure Interferometer (PdBI; \citealp{henshaw_2016b}). This comparison shows that a greater number of objects (28 versus 7) are detected in this new image due to the high angular resolution, sensitivity and dynamic range of ALMA.  

To characterize the physical properties of the compact cores, we use {\sc astrodendro} to first extract the sources.\footnote{The following parameters are used in computing the dendrogram: ${\tt min\_value}=~3\sigma_{\rm rms}$; ${\tt min\_delta}=~1\sigma_{\rm rms}$; ${\tt min\_npix}=60$.} The thick blue and red contours in the left-hand panel of Fig.~\ref{Figure:cont_map} show the locations of the 28 identified dendrogram leaves (the highest level in the dendrogram hierarchy, representing the smallest structures identified). Out of these leaves, four are rejected from further analysis as they reside at the very edge of the mosaic (thick red contours). 

\begin{table}
	\caption{Dendrogram leaves: physical properties (see \S~\ref{Section:results}).  } \vspace{0.1cm}
	
	\centering  
	\tabcolsep=0.15cm \small{
	\begin{tabular}{ c  c   c  c  c  c  c  }
	\hline
	RA$^{a}$ & 
	Dec$^{a}$ & 
	$R_{\rm eq}$$^{b}$ &
	$N_{\rm H, c}$$^{c}$  & 
	$M_{\rm c}$$^{d}$  & 
	$n_{\rm H, c, eq}$$^{e}$  & 
	$t_{\rm ff, c}$$^{f}$  
	\\ [0.5ex]

	&  
	& 
	&
	$\times10^{23}$ & 
	& 
	$\times10^{6}$ & 
	$\times10^{4}$ 
	\\ [0.5ex]
	
	& 
	& 
	(pc) &
	(cm$^{-2}$) & 
	(\solar) & 
	(cm$^{-3}$) & 
	(yr)
	\\ [0.5ex]
	
	\hline
	
18:57:08.09 &  2:10:28.05 &   0.026 &  8.3 &  3.1 &   1.2 &   4.0 \\ [0.5ex]
18:57:08.43 &  2:10:31.65 &   0.019 &  2.1 &  1.0 &   1.0 &   4.3 \\ [0.5ex]
18:57:08.06 &  2:10:33.00 &   0.013 &  1.9 &  0.4 &   1.4 &   3.7 \\ [0.5ex]
18:57:07.62 &  2:10:36.15 &   0.011 &  1.7 &  0.3 &   1.5 &   3.5 \\ [0.5ex]
18:57:09.13 &  2:10:36.60 &   0.013 &  2.5 &  0.5 &   1.5 &   3.5 \\ [0.5ex]
18:57:09.50 &  2:10:38.10 &   0.017 &  2.0 &  0.7 &   1.0 &   4.4 \\ [0.5ex]
18:57:08.00 &  2:10:39.30 &   0.031 & 10.6 &  7.7 &   1.8 &   3.3 \\ [0.5ex]
18:57:09.53 &  2:10:40.80 &   0.021 &  7.7 &  2.2 &   1.5 &   3.6 \\ [0.5ex]
18:57:09.23 &  2:10:41.70 &   0.026 &  2.4 &  1.9 &   0.7 &   5.1 \\ [0.5ex]
18:57:07.61 &  2:10:43.95 &   0.023 &  4.4 &  2.3 &   1.3 &   3.8 \\ [0.5ex]
18:57:08.46 &  2:10:44.40 &   0.014 &  2.1 &  0.6 &   1.6 &   3.5 \\ [0.5ex]
18:57:08.88 &  2:10:44.55 &   0.012 &  3.4 &  0.6 &   2.7 &   2.6 \\ [0.5ex]
18:57:07.91 &  2:10:45.00 &   0.024 & 13.9 &  5.6 &   2.9 &   2.6 \\ [0.5ex]
18:57:08.30 &  2:10:46.20 &   0.010 &  2.9 &  0.4 &   3.2 &   2.4 \\ [0.5ex]
18:57:08.67 &  2:10:48.60 &   0.019 &  3.1 &  1.6 &   1.6 &   3.5 \\ [0.5ex]
18:57:08.31 &  2:10:51.00 &   0.014 &  3.9 &  1.2 &   3.1 &   2.5 \\ [0.5ex]
18:57:07.62 &  2:10:51.75 &   0.021 &  2.5 &  1.5 &   1.1 &   4.2 \\ [0.5ex]
18:57:08.10 &  2:10:52.50 &   0.018 & 21.5 &  4.4 &   5.3 &   1.9 \\ [0.5ex]
18:57:08.50 &  2:10:53.55 &   0.026 & 28.3 & 10.4 &   4.1 &   2.1 \\ [0.5ex]
18:57:07.91 &  2:10:56.55 &   0.011 &  1.7 &  0.3 &   1.5 &   3.5 \\ [0.5ex]
18:57:08.37 &  2:10:57.60 &   0.014 &  6.3 &  1.5 &   4.2 &   2.1 \\ [0.5ex]
18:57:07.62 &  2:10:59.70 &   0.027 & 26.2 &  6.7 &   2.2 &   2.9 \\ [0.5ex]
18:57:08.16 &  2:11:05.25 &   0.016 &  7.5 &  1.6 &   2.5 &   2.8 \\ [0.5ex]
18:57:08.26 &  2:11:06.45 &   0.011 &  2.4 &  0.4 &   2.2 &   2.9 \\ [0.5ex]

	\hline
	\end{tabular}
	\vspace{0.2cm}

\begin{minipage}{0.5\textwidth}\footnotesize{
$^{a}$ Position of peak emission. \\
$^{b}$ Equivalent radius, $R_{\rm eq} = (N_{\rm pix}A_{\rm pix}/\pi)^{1/2}$. \\
$^{c}$ Peak beam-averaged column density. Uncertainty: $\sigma N_{\rm H, c}\sim50$ per cent. \\
$^{d}$ Leaf mass. Uncertainty: $\sigma M_{\rm c}\sim60$ per cent. \\
$^{e}$ Leaf number density. Uncertainty: $\sigma n_{\rm H, c, eq}\sim75$ per cent.\\
$^{f}$ Leaf free-fall time. Uncertainty: $\sigma t_{\rm ff, c}\sim40$ per cent.
}
\end{minipage}
}
\label{Table:leaf_info}
\end{table}

The derived equivalent radii of the leaves range between $0.01\,{\rm pc}<R_{\rm eq}<0.03\,{\rm pc}$, with a mean value of $\sim0.02\,{\rm pc}$ ($R_{\rm eq}\equiv~N_{\rm pix}A_{\rm pix}/\pi$, where $N_{\rm pix}$ is the total number of pixels and $A_{\rm pix}$ is the area of a single $0.15\,{\rm arcsec}\times0.15\,{\rm arcsec}$ pixel). The mean angular separation of the leaves is $\sim4.6\,{\rm arcsec}$, which corresponds to a projected linear separation of $\sim$0.065 pc at the distance of \irdc. Leaf peak beam-averaged column densities and masses are estimated following the assumptions listed above (subscript `c' is used to distinguish the core properties from the filament properties discussed above). Under these assumptions the derived peak beam-averaged column densities of the leaves range from $0.2\times10^{24}\,{\rm cm}^{-2}<N_{\rm H,c}<2.8\times10^{24}\,{\rm cm}^{-2}$. Leaf masses range from $0.3\,{\rm M_{\odot}}<M_{\rm c}<10.4\,{\rm M_{\odot}}$ (the uncertainty in our mass estimate is $\sim60$ per cent and our estimated mass sensitivity is $\sim0.2$\,\solar), which extends the distribution of detected core masses to lower values with respect to that derived by \citet[core masses ranging from $\sim$8 to 26$\,$M$_\odot$]{henshaw_2016b}. Estimating the mass within the boundaries of the leaves identified by \citet{henshaw_2016b} from our ALMA image, we note a $\sim30$ per cent reduction, on average (although this can be $\sim60$ per cent). This difference is likely due to the fact that the ALMA and PdBI images are sensitive to different spatial scales, and/or variation in the dust properties at the respective resolution of the two images. The equivalent number densities for each leaf, $n_{\rm H,c,eq}$ (i.e. the density of a structure with mass, $M_{\rm c}$, and radius, $R_{\rm eq}$, assuming spherical geometry), range between $0.7\times10^{6}\,{\rm cm}^{-3}<n_{\rm H, c, eq}<5.3\times10^{6}\,{\rm cm}^{-3}$, and the corresponding range in the estimated free-fall time is $1.9\times10^{4}\,{\rm yr}<t_{\rm ff,c}<5.1\times10^{4}\,{\rm yr}$. All derived values can be found in Table~\ref{Table:leaf_info}, where the leaves are listed in order of increasing declination.

\section{The early-stage anatomy of a protocluster hub}

Observational studies of molecular gas kinematics in both low-mass and high-mass star-forming regions have reported the presence of large-scale ($\sim2$\,pc) velocity gradients in hub-filament systems \citep{liu_2012, kirk_2013, peretto_2013, peretto_2014, tackenberg_2014}, possibly indicating mass transport (e.g. \citealp{gomez_2014}). In this scenario, protocluster clumps (analogous to ``beads-on-a-string'') grow in mass as material is accreted from parsec-scales along tributary filaments. Alternatively, recent single-dish and interferometric observations have revealed that, in some cases, these velocity `gradients' may instead be caused by the deceptive superposition of underlying substructure \citep[e.g.][]{hacar_2013, henshaw_2014}. In such instances, a centralized `hub' may represent the location where several sub-filaments, gathered together by large-scale motions, spatially overlap and merge (as inferred from e.g. the hydrodynamical simulations of \citealp{smith_2016}). Comparing to the `bead-on-a-string' analogy, where the protocluster hub is represented as large bead on a single string, each sub-filament in the latter scenario retains its integrity as an independent structure within the `hub' and possesses its own core population. 

The results presented in \S~\ref{Section:results} support the latter of these two scenarios. The anatomy of the H6 protocluster, at this early phase of its evolution, consists of several spatially-resolved intra-hub filaments, each with an associated population of embedded compact cores. This is consistent with both the kinematic and fragmentation analyses presented by \citet{henshaw_2014, henshaw_2016b}, respectively.

With widths of $\sim0.028\,\pm\,0.005$\,pc (\S~\ref{Section:results}), the H6 filaments are remarkably narrower than the proposed `quasi-universal' width of interstellar filaments ($\sim0.1$\,pc), a phenomenon which is supposedly independent of column density, evolutionary stage, and star formation efficiency \citep{arzoumanian_2011,federrath_2016}. Although there are several examples in the literature of filaments displaying a range of widths, rather than displaying a characteristic value (\citealp{pineda_2011, hennemann_2012, juvela_2012, schisano_2014}), it has been proposed that the width of filaments is intrinsically linked to their formation in shocks, and that this width may be set by the sonic scale of molecular cloud turbulence, $\lambda_{\rm sonic}$ \citep{arzoumanian_2011,federrath_2016}. This scale represents the transition from supersonic to subsonic turbulence and is given by \citep{federrath_2016}:
\begin{equation}
\lambda_{\rm sonic}=L\bigg[\frac{c_{\rm s}}{\sigma_{v, {\rm 3D}}}(1+\beta^{-1})\bigg]^{2},
\label{Eq:sonic}
\end{equation}
where $L$ is the cloud scale, $c_{\rm s}=(k_{\rm B}T/\mu m_{\rm H})^{1/2}$ is the isothermal sound speed at a gas temperature $T$, $\sigma_{v, {\rm 3D}}$ is the 3D velocity dispersion on the cloud scale, and $\beta=2c^{2}_{\rm s}/v^{2}_{\rm A}$ is the ratio of thermal to magnetic pressure ($v_{\rm A}=B/\sqrt{4\pi\rho_{\rm f}}$ is the Alfv\'{e}n velocity). 

By excluding the effects of magnetic fields from equation~\ref{Eq:sonic} (i.e. $\beta\rightarrow\infty$), we can estimate the corresponding sonic scale for the molecular gas in the H6 hub. Assuming that $T=13$\,K (see \S~\ref{Section:results}), $\sigma_{v, {\rm 3D}}\approx\sqrt{3}\sigma_{v, {\rm 1D}}\approx0.7$\,\kms \ (the mean value derived from line-fitting; \citealp{henshaw_2014}), and $L=0.35$\,pc (the projected extent of the filament in Fig.~\ref{Figure:cont_map}), we find a sonic scale $\lambda_{\rm sonic}\approx0.03$\,pc, in agreement with our derived filament width. Although the `cloud scale' (i.e. $L$) used above is somewhat dependant, in this simplistic analysis, on the field of view of our ALMA mosaic, this could imply that the observed intra-hub filaments have formed following the dissipation of turbulent energy via shocks. However, if the width of the H6 filament is indeed set by the sonic scale, then one would expect to observe a direct transition to coherence across the filament boundary similar to that reported by \citet{pineda_2010} in the Barnard 5 core in Perseus. This prospect will be investigated, along with a detailed description of the intra-hub kinematics in a future publication (Henshaw et al., in preparation). 

Our ALMA observations also reveal a number of cores associated with the sub-filaments, spanning a range of masses between $0.3\,{\rm M_{\odot}}<M_{\rm c}<10.4\,{\rm M_{\odot}}$. Previous studies of \irdc \ had reported a lack of cores in the mass range 2-8\,\solar \ \citep{henshaw_2016b}. A similar result was obtained by \citet{zhang_2015}, studying the P1 clump in IRDC G28.34+0.06, who reported a dearth of low-mass cores (by a factor of 5) in the 1-2\,\solar \ mass range. With only 24 cores identified in this study, we are unable to establish a robust statistical determination of the mass function in the H6 hub. However, with this new image, we identify a number of low-mass cores within the 2-8\,\solar \ mass range. We therefore conclude that the previously reported dearth of such low-mass objects may  be of observational, rather than physical origin, and that at least some low-mass cores are forming coevally in the immediate neighbourhood of more massive objects. The differences in the core mass distribution function between \irdc \ and G28.34+0.06 may be due to either an evolutionary effect (where the latter is at a more evolved phase in its evolution, as suggested by the several molecular outflows reported in the P1 clump) or, as discussed by \citet{zhang_2015}, because the lower mass cores in G28.34+0.06 are forming within the immediate vicinity of higher mass objects. In this latter scenario, the low mass objects may be indistinguishable from the extended emission of the P1 star-forming ridge. 

In summary, the high-angular resolution ($\sim0.01$\,pc) and high-sensitivity ($\sim0.2$\,mJy beam$^{-1}$) observations of the H6 clump have enabled us to investigate the anatomy of this complex protocluster hub with a high-degree of precision. Previously unresolved narrow ($0.028\,\pm\,0.005$\,pc) intra-hub filaments are revealed, as well as a number of low- and intermediate-mass cores. Although this study is insufficient to conclude that the detected narrow filaments (and their associated cores) are an intrinsic constituent of the initial phases of intermediate- and/or high-mass star formation, we have clearly demonstrated the need for high-angular resolution, high-sensitivity and high-dynamic range observations in determining the anatomy of such regions. Future investigations will focus on the complex gas dynamics of this region, with the aim of understanding how prestellar cores attain their mass during the early phases of evolution in protocluster hubs.
\section*{Acknowledgements}

We thank the anonymous referee for their comments and Nate Bastian for helpful discussions. This Letter makes use of the following ALMA data: ADS/JAO.ALMA\#2013.1.01035.S. ALMA is a partnership of ESO (representing its member states), NSF (USA) and NINS (Japan), together with NRC (Canada), NSC and ASIAA (Taiwan), and KASI (Republic of Korea), in cooperation with the Republic of Chile. The Joint ALMA Observatory is operated by ESO, AUI/NRAO and NAOJ. IJ-S. acknowledges the financial support received from the STFC through an Ernest Rutherford Fellowship (proposal number ST/L004801/1). PC and JEP acknowledge support from European Research Council (ERC; project PALs 320620). JCT acknowledges NASA grant ADAP10-0110. This research has benefited from the {\sc{astropy}} \citep{astropy_2013}, {\sc{matplotlib}} \citep{hunter_2007}, and {\sc{astrodendro}} (\url{www.dendrograms.org}) software packages.

%%%%%%%%%%%%%%%%%%%%%%%%%%%%%%%%%%%%%%%%%%%%%%%%%%

%%%%%%%%%%%%%%%%%%%% REFERENCES %%%%%%%%%%%%%%%%%%

% The best way to enter references is to use BibTeX:

\bibliographystyle{mnras}
\bibliography{References/references} % if your bibtex file is called example.bib

% Don't change these lines
\bsp	% typesetting comment
\label{lastpage}
\end{document}